\documentclass[useAMS,usenatbib]{mn2e}

\usepackage[dvips]{graphicx}
\usepackage[]{txfonts}
\def\simless{\mathbin{\lower 3pt\hbox
{$\rlap{\raise 5pt\hbox{$\char'074$}}\mathchar"7218$}}}   
\def\simmore{\mathbin{\lower 3pt\hbox
{$\rlap{\raise 5pt\hbox{$\char'076$}}\mathchar"7218$}}}   
\newcommand{\be}{\begin{equation}}
\newcommand{\ee}{\end{equation}}
\title[Peak energy of dissipative GRB photospheres]{The peak energy of dissipative GRB photospheres}
\author[Dimitrios Giannios]
{Dimitrios Giannios\thanks{E-mail: giannios@astro.princeton.edu}\\
Department of Astrophysical Sciences, Peyton Hall, Princeton
  University, Princeton, NJ 08544, USA}

\begin{document}
\date{Received / Accepted}
\pagerange{\pageref{firstpage}--\pageref{lastpage}} \pubyear{2011}

\maketitle

\label{firstpage}

\begin{abstract}
The radiation released at the transparency radius of an ultrarelativistic flow
can account for the observed properties of gamma-ray bursts (GRBs) 
provided that sufficient energy is dissipated in the sub-photospheric region.  
Here, I investigate how the peak energy of the $E f(E)$ spectrum
and its overall shape depend on the properties of the jet for various ``dissipative photospheres''.
I find that continuous energy release which results in electron
heating over a wide range of distances may be the key to explain the GRB emission. In this picture,
the peak of the spectrum forms at a Thomson optical depth of several
tens. The peak depends mainly on the bulk Lorentz factor $\Gamma$ of the flow and can, therefore, be
used to determine it. The $\Gamma$ is predicted to range from $\sim 10$ to 1000 from X-ray flashes 
to the brightest observed GRBs in agreement with recent observational
inferences. The Amati relation can be understood if the brightest
bursts are the least baryon loaded ones. Implications from this interpretation of the GRB emission
 for the central engine are discussed.
\end{abstract} 
  

\section{Introduction} 
\label{intro}

Although several thousand gamma-ray bursts (GRBs) have been observed so far, 
the mechanisms responsible for the GRB emission remain elusive.  Both synchrotron and 
photospheric models have been widely explored in the literature to explain the 
characteristic ``Band-like'' GRB spectrum which is characterized by smoothly 
connected power-laws (Band et al. 1993). The synchrotron models rely on a power-law
electron distribution accelerated at large distance (or small optical depths) in the jet. 
The photospheric models focus, instead, on radiation that is released when the jet becomes transparent. 

Photospheric models are attractive interpretation because they can result in high radiative efficiency
and naturally predict peak energies $E_{\rm peak}\sim1$ MeV close to
the observed values (Goodman 1986; Thompson 1994; M\'esz\'aros \& Rees 2000). 
The value of $E_{\rm peak}$ in this interpretation is coupled to the main
properties of the flow (e.g. luminosity $L$, and Lorentz factor $\Gamma$). 
Indeed, observations indicate that the peak of the $E f(E)$ spectrum
tracks the instantaneous gamma-ray luminosity and integrated energy 
during a burst and among different bursts, respectively (e.g., Amati et al. 2002; Yonetoku
et al. 2004; Ghirlanda, Nava \& Ghisellini 2010).

On the other hand, the energy dissipated in the base of the jet effectively 
thermalizes, so in the absence of additional dissipation at modest 
optical depths (i.e. further out) the emission from the transparency radius is quasi-thermal 
in sharp contrast to that typically observed. Significant dissipation of energy of some sort is 
required close to the photosphere of the flow to lead to the observed, 
smoothly-connected power-law spectra. The source of such dissipation 
may be (strong or many, weak) shocks (Rees \& M\'esz\'aros 1994; 
Lazzati \& Begelman 2011), magnetic reconnection (Giannios 2006), 
or nuclear collisions (Beloborodov 2010). 

Adopting the reconnection model for GRBs of Drenkhahn (2002), 
I have shown that magnetic dissipation leads to powerful photospheric emission
(Giannios 2006). The observed $\sim$MeV peak of the spectrum forms at
Thomson optical depth of several tens where radiation and electrons drop out of thermal 
equilibrium; the electrons turn hotter further out in the flow. 
Inverse Compton scattering at larger distance (smaller optical depth)
leads to the high-energy tail that can extend well into 
the GeV range. A Band-like spectrum naturally forms in this scenario.

Here, I develop a generic dissipative photospheric model applicable to
arbitrary dissipative process that results in electron heating. 
Both a flat rate of dissipation of energy 
(e.g. constant luminosity dissipated per decade of distance)
and localized dissipation events are explored (Sections 2, 3). 
The model predicts a relation of the peak $E_{\rm peak}$ of the observed 
emission with the properties of the flow; 
most sensitively depending on the bulk Lorentz factor 
$\Gamma$ (Section 4). Using the observed $E_{\rm peak}-L_{\gamma}$ 
relation I make inferences for the central engine of GRBs (Section 4.1).
Section 5 clarifies various aspects of the GRB variability in the context
of the model. Discussion and conclusions are presented in Section 6.

\section{Dissipative Photospheres}
\label{sec:disruption}

Energy dissipated at the very inner parts of the jet flow quickly thermalizes.
A substantial thermal component can also be built in an initially rather 
cold, magnetically dominated flow when magnetic energy is dissipated  
at large Thomson optical depth $\tau\gg 1$, e.g. as expected in a striped wind 
model (Drenkhahn \& Spruit 2002). The thermal luminosity $L_{\rm th}$ 
(dominated by radiation) is released at the 
transparency radius (defined as the distance at which $\tau=1$).
For typical parameters of the jet flow, the resulting quasi-thermal emission
peaks at  $E_{\rm peak}\sim 3k_{\rm B}T_{\rm obs}\sim 0.1-1$ MeV
(Goodman 1986, Thompson 1994; Daigne \& Mochkovitch 2002; 
Beloborodov 2010), very close to the typically observed values (Band
1993). In the absence of dissipation of energy close to the
photosphere, however, the emerging emission cannot account for 
the observed GRB spectrum. Though isolated cases 
for a strong quasi-thermal component in the GRB emission 
have been made (Ryde 2005; Ryde et al. 2010; Ryde et al. 2011), 
{\it the GRB spectrum generically has non-thermal appearance}.

The photospheric emission is, however, modified when additional 
energy release takes place close to the transparency radius.
It turns out (see next Section) that continuous electron heating
at a range of optical depths from $\tau\sim$several tens out to
 $\sim$0.1 may be the key to reproduce the observed emission. Continuous dissipation
results in a well defined distance where radiation and particles drop
out of equilibrium, the so-called {\it equilibrium radius}. 
The peak of the emission spectrum is determined
by the plasma temperature at this distance. 
Below we develop a general framework to calculate
the peak of the spectrum as function of the properties of the flow.

\subsection{A generic model}
\label{sec:jet}

Consider a jet coasting with bulk Lorentz factor $\Gamma$, total 
isotropic equivalent luminosity $L$ and baryon
loading $\eta\equiv L/\dot{M}c^2\simmore \Gamma$. In the presence of energetic particles
injected by the dissipative process, the flow can be loaded with
a modest number of pairs
(Pe'er et al. 2006; Vurm et al. 2011). Assuming 
$f_{\pm}$ electron+positron pairs per proton (i.e. $f_\pm =1$ for
$e-p$ plasma; hereafter pairs are referred to as electrons\footnote{
For a photospheric GRB model where the flow does not contain baryons
see Ioka et al. (2011).}), the rest-frame electron number density is 
$n_{\rm e}=f_{\pm} L/4\pi r^2\eta\Gamma
 m_{\rm p}c^3$. Clearly both $\Gamma$ and $f_{\pm}$ can vary with
distance (because of acceleration of the flow and energetic particle
injection that result in pair creation, respectively).
Here, I treat these quantities as constants with the main focus been 
on their values close to the equilibrium radius.  
The Thomson optical depth $\tau\equiv n_{\rm e}\sigma_{\rm T}r/\Gamma$
as function of distance is
\be
\tau=37\frac{L_{53}f_{\pm}}{r_{11}\eta_{2.5}\Gamma_{2.5}^2},
\ee
where all quantities are in cgs units and the $A=A_x10^{x}$ notation
is adopted. Setting $\tau=1$, the Thomson photosphere is located at
\be
r_{\rm ph,11}=37\frac{L_{53}f_{\pm}}{\eta_{2.5}\Gamma_{2.5}^2}.
\ee 

We consider a flow which carries a thermal component
of luminosity $L_{\rm th}$ that is a substantial fraction $\epsilon$ of that of the flow:
$L_{\rm th}=(4/3)4\pi r^2\Gamma^2aT_{\rm th}^4c=\epsilon L$. 
This can be realized in both fireballs and dissipative, magnetically 
dominated flows (e.g. Drenkhahn 2002)
The (rest frame) plasma temperature is
\be
k_{\rm B}T_{\rm th}=1.1\frac{\epsilon^{1/4}L_{53}^{1/4}}{r_{11}^{1/2}\Gamma_{2.5}^{1/2}}\quad \rm keV.
\ee
Inspection of eqs. (2) and (3) reveals that
the plasma typically cools to sub-keV temperature close to the photosphere.
It turns out, however, that the dissipative process heats up the flow 
to a temperature well in excess of $T_{\rm th}$ before it reaches the photosphere. 

Suppose that a dissipative process injects energy in the flow heating the electrons
(see Section 3 for discussion on the physical justification of such assumption).
I consider two different situations for the radial dependence
of the dissipation rate: i) a gradual rate of energy release of rather 
flat profile (i.e. constant rate of energy dissipated per e-folding of distance):
$dL_{\rm d}/dr=L_{\rm d,o}/r$ and ii) dissipation that is localized to a narrow range
in distance. 

In the case of the gradual energy release, the heating rate per unit
volume in the rest frame of the flow is 
$P_{\rm h}=L_{\rm d,o}/4\pi \Gamma r^3$. If the thermal component
$L_{\rm th}$ is built by the dissipation of energy from smaller distances
(and including adiabatic cooling of the photons) 
one gets $L_{\rm th}=(3/2)L_{\rm d,o}$. Allowing for a comparable amount
of additional heating even deeper in the flow, we set $L_{\rm d,o}=L_{\rm th}/3=\epsilon
L/3$.

Heating of the electrons is balanced by radiative cooling\footnote{Adiabatic
cooling can be shown to be negligible for the electrons.}.
Electrons and photons are in thermal equilibrium 
at large depth with their common temperature given by
eq.~(3)\footnote{In reality, the electrons maintain a slightly higher
temperature from radiation because of external heating.}. 
This equilibrium is broken once heating and cooling balance  
leads to $T_{\rm e}>T_{\rm th}$. It can be shown (see Giannios 2006) that Compton cooling
$P_{\rm IC}=4n_{\rm e}\Theta_{\rm e}c\sigma_{\rm T}U_{\rm r}$ is the dominant 
cooling mechanism for the electrons in this region ($\Theta_{\rm
  e}=k_{\rm B}T_{\rm e}/m_{\rm e}c^2$ and $U_{\rm r}$ is the energy
density of radiation). 
Equating dissipative heating and cooling ($P_{\rm h}=P_{\rm IC}$)
and setting $U_{\rm r}=aT_{\rm th}^4$ one derives the electron temperature as 
function of distance 
\be
k_{\rm B}T_{\rm e}=1.5\frac{r_{11}\Gamma_{2.5}^2\eta_{2.5}}{f_{\pm}L_{53}}\quad \rm keV.
\ee 
The location where radiation and electrons drop out of equilibrium is
found by setting $T_{\rm th}=T_{\rm e}$, defining the (maximum) {\it equilibrium radius} $r_{\rm eq}$:
\be
r_{\rm eq,11}=0.80\frac{L_{53}^{5/6}\epsilon^{1/6} f_{\pm}^{2/3}}{\Gamma_{2.5}^{5/3}\eta_{2.5}^{2/3}}.
\ee
 
At the equilibrium radius $r_{\rm eq}$, the Thomson optical depth and the temperature of the electrons are, respectively
\be
\tau_{\rm eq}=46\frac{L_{53}^{1/6} f_{\pm}^{1/3}}{\epsilon^{1/6}\Gamma_{2.5}^{1/3}\eta_{2.5}^{1/3}}
\ee
and 
\be
k_{\rm B}T_{\rm eq}=1.2\frac {\epsilon^{1/6}\Gamma_{2.5}^{1/3}\eta_{2.5}^{1/3}}{L_{53}^{1/6} f_{\pm}^{1/3}}\quad \rm keV.
\ee

Note that the optical depth and temperature of the plasma at the equilibrium 
radius depend very weakly on the flow parameters and are in the range of
several tens and $\sim 1$ keV, respectively.
The equilibrium radius is of particular importance since it is the location
at which the peak of the spectrum forms 
under a wide range of conditions. The observed peak of the spectrum is
\be
E_{\rm peak}\simeq \frac{4}{3}3\Gamma k_{\rm B}T_{\rm eq}\simeq 1.5 
\frac {\epsilon^{1/6}\Gamma_{2.5}^{4/3}\eta_{2.5}^{1/3}}{L_{53}^{1/6} f_{\pm}^{1/3}}\quad \rm MeV.
\ee
At distance $r>r_{\rm eq}$, and using eqs. (1) and (4), 
the Compton $y$ parameter\footnote{For relativistically expanding
fluid the number of scatterings scales as $\sim \tau$ and not
$\propto \tau^2$ as in a static medium (see Giannios 2006).} is found
to be $y=4\Theta_{\rm e}\tau \sim 0.4$ 
{\it independently} of distance or the parameters of the flow leading
to significant up-scattering of the photons. 
The Compton parameter is close to unity because the incoming
radiative luminosity $L_{\rm th}$ and dissipation rate $L_{\rm d,o}$ are comparable. 
Up-scattering of the radiation emerging from $r_{\rm eq}$
at larger distance leads to broader spectra and to a flat high-energy tail
above $E_{\rm peak}$.  The high-energy tail is followed by an exponential
cutoff at the energy that corresponds to the electron temperature
 at the distance where the dissipation stops (see next section).

\subsection{Numerical Results}

The analytical arguments presented in the previous Section are fully supported by detailed Monte-Carlo
radiative transfer simulations. The code developed in Giannios (2006) is used to
study the electromagnetic spectrum emerging from a dissipative
photosphere. The inner boundary of the calculation is set at the
equilibrium radius where photons following a thermal distribution
of temperature $T_{\rm eq}$ are injected. The photons
are followed throughout the photosphere until the flow reaches an
optical depth of $\tau=0.1$ where the outer boundary is set.    
The calculation includes Compton scattering, relativistic effects, while
the electron temperature is iterated until the heating
rate is matched by Compton cooling everywhere in the flow. 
We do not include synchrotron emission and 
non-thermal particle acceleration. While both are important in 
determining the exact spectrum, they are model dependent.
For this paper, however, it is important that the peak of the
emission is set at the distance where radiation and
matter drop out of equilibrium and rather independently of the details
of non-thermal processes. In the following, 
both extended dissipation with $dL_{\rm d}/dr=L_{\rm d,o}/r$
and localized dissipation are investigated.

\subsubsection{Extended Dissipation}

\begin{figure}
\resizebox{\hsize}{!}{\includegraphics[angle=270]{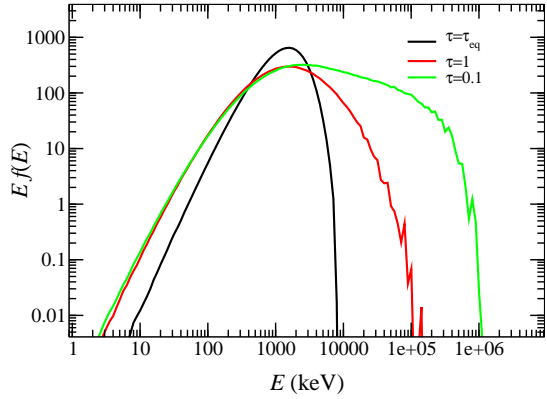}}
\caption[] {Numerically calculated spectra at different optical depths in the jet
for the reference values of the parameters and for extended dissipation of energy.
The thermal injected photon spectrum at the equilibrium radius (black line) evolves into
a broader spectrum at the $\tau=1$ surface (red line) and develops a 
flat power-law tail at the outer radius (corresponding to $\tau=0.1$; green line). 
The peak of the $E f(E)$ emerging spectrum is determined at the equilibrium radius.} 
\label{fig:diagram}
\end{figure}

Assuming extended dissipation of energy,
in Fig.~1 I plot the radiation spectrum at various distances of the flow
for the reference values of the parameters ($L=10^{53}$ erg$\cdot$s$^{-1}$, $\eta=\Gamma=300$,
$\epsilon=0.3$, $f_{\pm}=1$). Note that the thermal emission
at $r_{\rm eq}$ evolves into one of non-thermal appearance when passing through the $\tau=1$
surface building a flat high-energy tail. Inverse Compton scattering
with Compton $y$ parameter $y\simless 1$ results in a flat ($E f(E)\sim E^0$) emission above the peak.
The outer boundary of the calculation is set at larger distance 
that corresponds to $\tau=0.1$. The high-energy cutoff $E_{\rm cut}$ is determined by
the temperature of the electrons at the outer boundary (in this example
$k_{\rm B}T_{\rm e, out}\simeq 300$ keV resulting in $E_{\rm cut}\simeq \Gamma k_BT_{\rm e, out}\simeq 100$ 
MeV). While in this paper, I set the outer boundary of dissipation by hand at $\tau=0.1$, 
more detailed models, such as the magnetic reconnection model of
Drenkhahn \& Spruit (2002) predict where this cutoff takes place. 

In Fig.~2 I show the resulting spectra for different values of the parameters.
The peak of the emission $E_{\rm peak}$ is very close to the analytic expression (8). 
The overall shape of the spectra is very similar with only the peak and cutoff 
energies depending on the parameters. A Band-like spectrum is  
reproduced for very different parameters of the flow. The high-energy 
tail has spectral slope $f(E)\propto E^{\beta}$ with $\beta\simless -1$
as observed. Below the peak energy the slope, as measured in the 10-100 keV
range, is $f(E)\propto E^{\alpha}$ with $\alpha \sim 0-1$. The slope 
$\alpha$ is consistent but in the rather hard range in
comparison with the observed distribution. Note, however, that synchrotron and 
synchrotron-self-Compton emission from larger distances (not included
in this work) can soften the spectrum below the peak (Giannios 2008;
Vurm, Beloborodov \& Poutanen 2011).

\begin{figure}
\resizebox{\hsize}{!}{\includegraphics[angle=270]{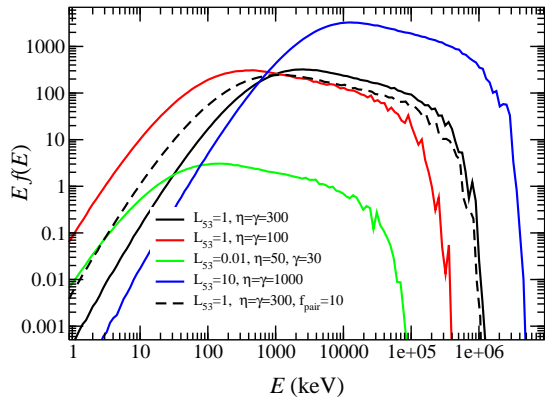}}
\caption[] {Emerging spectra for different values of the parameters and for 
extended dissipation of energy. The spectrum
has a Band-like shape practically independently of the luminosity, baryon loading
and Lorentz factor of the flow. The peak of the spectrum follows the scaling predicted by 
eq. (8). The high energy cutoff is set by the location of the outer boundary of the simulation.} 
\label{fig:diagram}
\end{figure}

\subsubsection{Localized Dissipation}

The process that dissipates energy in the jet may result in energy release over a narrow
range of distances. Here, I investigate how a thermal component carried
by the flow is modified depending on the optical depth at which such energy release takes place.

I assume that the thermal component $L_{\rm th}=\epsilon L$ is reprocessed by Compton 
up-scattering through a narrow region of hot electrons. The electrons
are heated at a rate $L_{\rm d,o}=L_{\rm th}$ over a range of Thomson
depth $\tau_{\rm diss}...\tau_{\rm diss}/2$
(i.e. a factor 2 in distance). The electron temperature is determined
by heating-cooling balance. Setting the various parameters to their reference values,
Fig.~3 shows the emerging spectrum for different values of $\tau_{\rm diss}$.       
Dissipation at large optical depths $\tau_{\rm diss}\simmore 10$
leads to a broadened, Planck-like distribution. The resulting 
narrow emission spectrum has a steep rise/decline below/above the
peak. Dissipation at $\tau_{\rm diss}\simless 1$ leads to a hot
region above the photosphere. Most of the photons do not interact with the hot electrons
leading to a distinct quasi-thermal component while; photons
that are up-scattered at least once form a high-energy component. 
Although occasionally GRB spectra show 
such multi-component behavior in the keV-MeV regime, this in not typical.  
For $\tau_{\rm diss}\sim 3$ the two components are smoothly connected and 
the resulting spectrum compares more favorable with observations.
For a model where the dissipation (through magnetic reconnection)
takes place at a narrow region of $\tau\simmore 1$ see Thompson (1994).

Summarizing, localized dissipation in general does not account for 
observations because it either leads to narrow or multi-component 
spectra in the $\sim$MeV energy range. 
There is a limited range of optical depths of $\tau_{\rm diss}\sim 3-5$
where dissipation results in a smooth Band-like spectrum.  

\begin{figure}
\resizebox{\hsize}{!}{\includegraphics[angle=270]{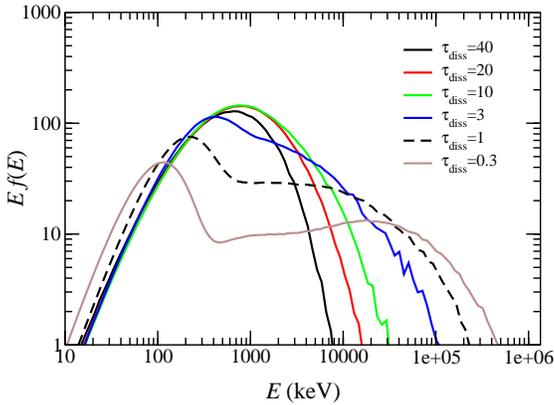}}
\caption[] {Emerging spectrum for dissipation of energy at a narrow range
of distance (or optical depth $\tau_{\rm diss}$) for different $\tau_{\rm diss}$. 
Dissipation at large optical depths $\tau_{\rm diss}\simmore 10$
leads to narrow emission spectrum. Dissipation at $\tau_{\rm diss}\simless 1$ leads to distinct
thermal and high-energy components. For $\tau_{\rm diss}\sim 3$ the two components are
smoothly connected.} 
\label{fig:diagram}
\end{figure}

\section{The dissipative mechanisms}

In the previous analysis we simplified the calculation 
assuming thermal electrons continuously heated 
by an external agent. Here we elaborate on the
possible sources for smooth ``volume" heating
of the GRB flow (Ghisellini \& Celotti 1999; 
Stern \& Poutanen 2004; Pe'er, M\'esz\'aros 
\& Rees 2006).

Models for gradual energy release that heats 
the electrons involve magnetic reconnection (Giannios 2006), 
multiple weak shocks (Ioka et al. 2007;
Lazzati \& Begelman 2010)\footnote{The, more often
invoked for the GRB emission, strong internal shocks 
(Rees \& M\'esz\'aros 1994) are unlikely to lead to 
smooth/gradual heating of the electrons. Such shocks, instead,
accelerate particles practically instantaneously to ultrarelativistic
energies leaving them to cool radiatively on a longer time scale
(see, however, Ramirez-Ruiz 2005).} 
and neutron-proton collisions (Beloborodov 2010).
In the first case, the energy is initially stored in the 
magnetic field while in the latter cases in relative 
bulk motions within the jet. In the following I argue
that, independently of the dissipative mechanism,  
a large fraction of the energy dissipated close to the
equilibrium radius is expected to be stored
to (mildly) relativistic protons. Coulomb $e-p$ collisions
effectively drive the energy from protons into the electrons
(see Beloborodov 2010).
The electrons then transfer their energy into the photon field
through Compton scattering. The characteristic 
electron equilibrium temperature is sub-relativistic 
and determined by heating-cooling balance.   

Multiple weak shocks in the jet or elastic neutron-proton collisions
(close to the distance where the neutron and proton fluids decouple)
can be expected to heat the protons at mildly relativistic 
temperatures with the heating maintained over a range
of distances in the jet. When the heating takes place at 
optical depth of tens, the Coulomb coupling of 
$\simless$ GeV protons with 
$\sim$keV electrons (see, e.g., Stepney 1983) can be shown to be effective in transferring 
the energy into the electrons within one expansion timescale of the jet.
These, slowly heated, electrons pass their energy into the photon field
effectively through inverse Compton scattering.
In addition to mildly relativistic protons, shocks and inelastic $n-p$
collisions can inject high-energy particles. Non-thermal 
processes can result in a modest pair loading in the jet
(Pe'er et al. 2006; Beloborodov 2010)\footnote{Note, however, 
that if the energy density of the magnetic field is higher than 
that of the radiation, ultrarelativistic electrons 
cool mainly through synchrotron emission and the pair creation is
further limited.} 
but not expected to directly affect the peak energy of the emission as long they 
are not energetically dominant.   

Magnetic reconnection can take place effectively 
at large optical depths (deep in the jet) through, e.g., tearing instabilities
of the current sheet (Loureiro, Schekochihin \& 
Cowley 2007; Uzdensky, Loureiro \& Schekochihin 2010) assisted/induced
by the acceleration of the jet (Lyubarsky 2011).\footnote
{A possible speed up of the reconnection rate at Thomson thin
 conditions (Lyutikov \& Blandford 2003; McKinney \& Uzdensky 2012) 
can have interesting implications but is not 
required in this picture.}
Magnetic reconnection takes place in multiple 
dissipative centers that accelerate particles, 
heat the plasma and drive fast bulk motions. 
The fast motions in the downstream of the reconnection
region are required for sufficiently fast reconnection.
The bulk motions are dissipated further downstream
in shocks resulting in hot protons. Coulomb collisions 
can effectively couple the proton energy to the electrons which,
in turn, couple it to the radiation field (as discussed above).  
In this strongly magnetized plasma ($U_{\rm B}\simmore U_{\rm ph}$),
sub-relativistic electrons can thermalize very efficiently through exchange of synchrotron 
photons (the so-called ``synchrotron boiler''; Ghisellini et al. 1998)
before they have the chance to cool through inverse Compton scattering.
An alternative picture where the energy released
in the reconnection process  drives MHD waves
(instead of bulk motions) is described in Thompson (1994).
Photons extract the energy from the waves through 
scattering on electrons (Compton drag). 
Also in this picture the electron velocity is limited to 
sub-relativistic speed and close to a Compton equilibrium 
with the radiation field. The analysis of the previous
section may apply to this scenario as well. 
 
In summary the details of the dissipation process 
(and the related non-thermal processes) may well
affect the pair injection rate and result in additional features
in the spectrum (e.g. powering a pair
annihilation line at $E\sim \Gamma m_{\rm e} c^2$)
but do not change the basic picture of 
sub-relativistic electrons strongly coupled to the
radiation field with their energy determined by a balance
of dissipative heating and radiative cooling. It is this basic
process at the equilibrium radius that determines the peak of the emerging  
emission. 

\section{Inferring the bulk Lorentz factor of the flow}

In this rather general framework of dissipative photospheres,
there is a close connection between the peak of the spectrum
and the properties of the flow summarized in eq.~(8). Eq.~(8)
can be solved in terms of $\Gamma$:
\be
\Gamma=280 E_{\rm peak,MeV}^{3/5}\epsilon_{-0.5}^{-1/10}L_{53}^{1/10}
f_{\pm}^{1/5}\big(\frac{\eta}{\Gamma}\Big)^{-1/5},
\ee
where I chose to express $\Gamma$ as function of the ratio $\eta/\Gamma\simmore 1$
instead of $\eta$. A clear implication from the last expression 
is that {\it $\Gamma$ depends mainly on the 
peak energy $E_{\rm peak}$ with extremely weak dependence on the 
physical properties of the jet}. To first approximation,
one can have a fair estimate of the Lorentz factor 
just from the observed peak of the GRB emission. 
From eq.~(9) it is clear that the model predicts that 
the high-peaked GRBs come from faster jets.
This is in sharp contrast to the synchrotron
internal shock model where the opposite holds
true ($E_{\rm peak}\propto \Gamma^{-2}$).

One can somewhat improve on the estimate
of the bulk $\Gamma$ of the jet from observables by assuming a 
gamma-ray luminosity $L_{\gamma}\sim \epsilon L$ 
and using the observed $L_{\gamma}$ and $\epsilon\sim 1$. 
This interpretation of the peak of the emission implies that 
the brightest GRBs with $E_{\rm peak}\sim$ several MeV
and $L_{53}\sim 10$ (e.g. Abdo et al. 2009)
come from the most relativistic $\Gamma \sim 1000$
flows while weaker X-ray flashes with $E_{\rm peak}\sim 30$ keV
and $L_{53}\sim 0.01$ come from  ``slower" 
jets of $\Gamma\sim 20$.  Low-luminosity GRBs
(e.g.  Soderberg et al. 2006) and the X-ray flares 
that follow many bursts (Burrows et al. 2005; 
Chincarini et al. 2010) can also be a result of a 
dissipative photosphere from yet slower jets of $\Gamma\simless 10$.
In particular low-luminosity GRBs may come
form $\Gamma\sim 3$ jets that result in only modest
relativistic beaming of their emission and may, therefore, account for the
larger observed local rate of these events 
(Soderberg et al. 2006) in comparison to classical GRBs. 
As discussed in Section 6, observational estimates of
the Lorentz factor of GRBs support the model prediction that higher
$E_{\rm peak}$ bursts are coming from higher $\Gamma$ flows.

\subsection{Additional inferences for the flow using observed
  correlations of the GRB emission}

It has been recognized for some time that various {\it time
integrated} quantities over the duration of a burst may
correlate with each-other, e.g., the peak energy $E_{\rm peak}$ with 
the isotropic gamma-ray energy $E_{\gamma}$ (Amati et al. 2002). 
Even in the presence of outliers (Band \& Preece 2005; Nakar \& Piran
2005), these correlations 
may teach us a lot about the GRB physics.
The emission in the photospheric models 
is, however, connected to the {\it instantaneous} properties of
the flow. A change in any of the properties 
(e.g. luminosity or baryon loading) during the burst shifts the location
of the photosphere and its appearance.  Ghirlanda 
et al. (2010; 2011) found that there is a time dependent 
Amati-like relation of $E_{\rm peak}(t)$ and $L_{\gamma}(t)$ during
the evolution of bursts where $E_{\rm peak} \simeq 1L_{\gamma,53}^{1/2}$ MeV.

If this relation is verified by more data, it implies for the photospheric models that
the Lorentz factor correlates with the luminosity of the flow.
In this interpretation, one can derive an estimate of the Lorentz factor 
of the flow directly from the observed gamma-ray luminosity
using eq. (9), the instantaneous $E_{\rm peak}-L_{\gamma}$ correlation
and that $L_{\gamma}\sim L_{\rm th}=\epsilon L$:
\be
\Gamma=310 E_{\rm peak,MeV}^{4/5}\epsilon_{-0.5}^{-1/5}f_{\pm}^{1/5}\Big (\frac{\eta}{\Gamma}\Big)^{-1/5}.
\label{gammati}
\ee
Note that the Lorentz factor depends sensitively on a single
observable, namely, the peak energy. Therefore, $E_{\rm peak}$ 
can be use the infer the Lorentz factor.
In terms of properties of the flow, I find that $\Gamma= 200 \epsilon_{-0.5}^{1/5}L_{53}^{2/5}
(\eta/\Gamma)^{-1/5}f_{\pm}^{1/5}$, i.e., that 
more luminous bursts are less baryon loaded
with $\Gamma\propto L^{2/5}$. 
Note, however, that a systematic dependence of any other quantity
with, say, the luminosity of the flow can distort this $\Gamma-L$ scaling.
Finally, the $\Gamma-L$ relation can be combined with eq.~(7) to find
the temperature of the flow at the equilibrium radius:
\be
k_{\rm B}T_{\rm eq}\simeq 0.8  \epsilon_{-0.5}^{3/10}L_{53}^{1/10} f_{\pm}^{-1/5}\Big (\frac{\eta}{\Gamma}\Big)^{1/5}\rm \quad keV.   
\ee
Note that the temperature of the flow clusters in the $\sim$keV range
and has very weak dependence on the parameters of the jet. As
discussed in Section 6 such clustering of the peak of the
emission in the rest frame of the flow has been observed.   

The instantaneous $E_{\rm peak}-L_{\gamma}$ relation also puts interesting constraints on the distance 
where the peak of the spectrum forms. Using the expression (5) for
$r_{\rm eq}$ and eq.~(\ref{gammati}),  I arrive to
\be
r_{\rm eq}=2.1\times 10^{11} L_{53}^{-1/10}\epsilon_{-0.5}^{-3/10}f_{\pm}^{1/5}\Big (\frac{\eta}{\Gamma}\Big)^{-1/5}\quad \rm cm.
\ee    
The equilibrium distance $r_{\rm eq}$ depends very weakly on the various parameters. Even allowing
for orders of magnitude variations in the luminosity  and (as expected) more modest changes
in other parameters from burst to burst (or during the evolution of a burst), $r_{\rm eq}$
varies at most by a factor of a few.

\section{GRB variability}

GRBs are variable on timescales as short as milliseconds. 
Their lightcurves are characterized by multiple pulses 
(e.g. Fenimore et al. 1995) that typically last for seconds and
 show characteristic spectrum that often 
evolves from hard to soft (e.g. Hakkila \& Preece 2011) and/or tracks the 
instantaneous flux of the flow (e.g. Ghirlanda et al. 2011). Can the photospheric
model account for the temporal GRB behaviour?

The small radius of emission in the photospheric models
allows for fast variability down to $t_{\rm v}\sim r_{\rm
  ph}/2\Gamma^2c \sim $sub-msec timescales (depending
on the parameters of the flow; see also Giannios \& Spruit 2007). 
Allowing the continuous heating out to optical depth $\tau \sim
0.1$ ($r_{\rm out}= 10 r_{\rm ph}$) can lead to the 
multi MeV high-energy tail delayed by $ r_{\rm
  out}/2\Gamma^2c\sim$several msec with respect to 
the MeV peak (with the GeV emission potentially more delayed depending
on the distance at which it takes place). The steady-state jet model developed here 
is not applicable to study extremely short (sub-msec) timescales
(for which the steady-state assumption breaks down) but is well suited to study
the evolution of the burst on longer timescales. 
In particular, the $\sim$sec duration GRB pulses can be accurately studied
as a sequence of steady state models where the properties of the jet 
($L=L(t)$, $\Gamma=\Gamma(t)$, etc)
vary on this timescale. The observed variability on second
timescales in the well studied $\sim 10-1000$ keV energy range 
is likely to be dominated by temporal evolution of the properties 
of the jet rather than delays introduced by propagation effects of the
ejecta.

The spectral evolution during a GRB pulse depends on how 
$\Gamma(t)$ and $L(t)$ evolve during the pulse. 
Giannios \& Spruit (2007) have shown that {\it If}, for
instance, $\Gamma(t)$ has a positive power-law dependence on $L(t)$ 
during individual pulses then the peak of the spectrum tracks the observed flux (as seen
observationally in Ghirlanda et al. 2011). In this case, while the 
flux declines after the peak luminosity of a pulse, the peak of the emission
spectrum also declines and the spectrum softens. As a result the 
pulse duration on the softer X-ray bands is longer than in the harder ones. 

In summary, continuous dissipation close to the photosphere can 
allow for fast evolving (sub-msec) emission. Furthermore, it is possible to account 
for observed temporal properties of GRBs given specific assumptions
about the behaviour of the central engine. The reasons for which the 
engine operates in such a fashion is, however, not addressed by this work. 

\section{Discussion and Conclusions}
\label{sec:discussion}

In this paper, I have explored a wide range of 
``dissipative photosphere'' models to pin
down the location where the peak $E_{\rm peak}$
of the spectrum forms and how it connects to the properties 
of the jet. The resulting expression (9) of my analysis
summarizes this connection, is quite general and 
rather independent of the physical model for energy 
dissipation.

I have explored both localized and continuous in distance dissipation
of energy. Localized heating of electrons over a narrow 
range of distance is shown here to have difficulties to 
account for observations. If the dissipation takes place
at large optical depth, the emerging spectrum is rather
narrow, while dissipation at $\tau\simless 1$ has two
distinct components in the $\sim$MeV energy range, both
in conflict to the majority of observed bursts. However,
localized dissipation that takes place
at optical depth $\tau\sim 3$ results is a more promising
spectrum. Thompson (1994) discusses a scenario where 
such dissipation is possible.   

I find that {\it continuous} dissipation
of energy over a wide range of optical depths 
can naturally give the Band-like spectrum with 
peak and slopes above and below the peak at the observed range.
Magnetic reconnection in a jet that contains
small-scale field reversals naturally results in such flat dissipation
profile throughout the photospheric region. Alternatives
such as multiple, weak shocks or neutron-proton collisions
can also result in continuous energy injection. 
The MeV peak of the spectrum forms at
the distance where radiation and the electrons
drop out of thermal equilibrium. In the context of this model
the peak energy is mainly determined by the 
bulk Lorentz factor of the flow. The observed $E_{\rm peak}$
can, therefore, be used to infer $\Gamma$. 

The model predicts that the peak energy $E_{\rm peak}$ positively correlates
with $\Gamma$. I find that weak 
GRBs (the spectrum of which peaks in the X-ray regime; the so called 
X-ray flashes) and X-ray flares that follow the bursts 
potentially come from $\Gamma\sim 10$ jets while the
brightest observed {\it Fermi-LAT} bursts, with peak energy at several
MeV, come from the fastest $\Gamma\sim1000$ flows.
Recently (Ghirlanda et al. 2012) used the peak time of the afterglow
emission to estimate the bulk Lorentz factor $\Gamma$ of 31 GRBs.
They verified that brighter bursts are characterized by higher
$\Gamma$ and showed that the peak energy of the emission at the rest frame
of the jet clusters at several keV; both findings are unique predictions of
this work (see eqs. 7, 11).  

The observed relation of the peak of the spectrum with the {\it
  instantaneous} burst luminosity 
indicates a close coupling of the emerging spectrum
to instantaneous properties of the flow  (Ghirlanda et al. 2011). 
It has already been pointed out in Giannios \& Spruit (2007) that
the $E_{\rm peak}-L_{\gamma}$ relation implies, 
in the context of the reconnection model, that
brighter segments of a burst come from higher $\Gamma$ 
(i.e., ``the brighter the cleaner'') jet. 
I come to the same conclusion for the generic photospheric 
model developed here where the bulk Lorentz factor $\Gamma$
and the luminosity of the flow $L$ scale as, roughly, $\Gamma\propto L^{2/5}$.
Note again that independent constraints coming from afterglow modeling
indicate a similar correlation for $\Gamma$ 
and the gamma-ray energy $E_{\gamma}$ (Liang et al. 2010).

In this model, the MeV peak forms in a 
fairly compact region, while the high-energy tail 
forms at a larger 
distance. The tail can extend to multi-GeV energy without
suffering attenuation due to pair creation and can exhibit delays of
the order of $r_{\rm GeV}/\Gamma^2c$ with respect to
the MeV emission that may 
be of order of seconds (as observed in some {\it Fermi-LAT}
bursts, see Abdo et al. 2009; $r_{\rm GeV}$ 
is the radius where GeV emission takes place).
While a different mechanism may be invoked 
for the dissipation at large distance that leads to the GeV emission
(M\'esz\'aros \& Rees 2011), the continuous spectrum from MeV to GeV 
energy indicates a single dissipative mechanism\footnote
{but not necessarily a single emitting region
for the MeV and GeV emission.} (see, e.g., Bosnjak \& Kumar 
2012). 

Interestingly, using the observed instantaneous $E_{\rm
  peak}-L_{\gamma}$ relation, the equilibrium distance $r_{\rm eq}$ 
(where the peak of the emission is set) can be shown to depend extremely weakly 
on the parameters of the flow with $r_{\rm eq}\simeq 2\times 10^{11}$ cm.
This value is similar to the radii of Wolf-Rayet stars,
probable progenitors of long-duration GRBs. This could indicate that  
interactions of the jet with the progenitor are important for 
energy dissipation (e.g. Lazzati \& Begelman 2010). 
On the other hand, a similar correlation is observed in 
short-duration GRBs, supporting the idea
of a progenitor-independent dissipative mechanism
(Ghirlanda et al. 2011). 

The inferred ``the brighter the cleaner" property of GRB 
flows has profound implications for the nature of the central engine.
Both accreting black holes and rapidly rotating magnetars
have been invoked for launching the jet.
Models that invoke accretion into a black hole (Woosley 1993)
are not sufficiently developed to predict the amount of baryons
that make it into the jet. In the millisecond magnetar model
for GRBs (Usov 1992; Metzger et al. 2011), the calculation of the
baryon loading is more tractable. The magnetars born with the 
strongest fields drive the
brightest bursts and also give, averaged over the GRB
duration, less baryon loaded flows. In particular, oblique rotators result in
$\Gamma \propto L^{0.6}$ (but with a large scatter; see 
Metzger et al. 2011). This is a rather intriguing
result since it might point to a rather complete picture for GRB
physics (which is, however, hardly unique). 
A central engine consists of a protomagnetar
(with its magnetic field axis, in general, misaligned to the rotational axis),
that gives rise to a magnetically dominated jet that contains
field reversals on small scale (striped wind). 
Magnetic reconnection proceeds in the jet at a wide range of
distances from the central engine and at a rather flat rate (Drenkhahn
\& Spruit 2002). The bright MeV emission of the GRB emerges from the Thomson 
photosphere of the flow while residual dissipation can extend
the high-energy tail above the MeV peak well into the GeV regime (Giannios 2008).

\section*{Acknowledgments}
DG acknowledges support from the Lyman Spitzer, Jr. Fellowship awarded 
by the Department of Astrophysical Sciences at Princeton University
and from the Fermi 4 Cycle grant number 041305.

\end{document}